\documentclass[aps,prr,twocolumn,preprintnumbers,amsmath,amssymb,superscriptaddress,10pt]{revtex4-2}
\usepackage{amsmath,graphicx}
\usepackage[utf8]{inputenc}
\usepackage[T1]{fontenc}
\usepackage{xcolor}
\usepackage{textcomp}
\usepackage{bm}
\usepackage{slashed}
\usepackage{physics}
\usepackage{amsmath} 
\usepackage{tikz}
\usepackage{mathdots}
\usepackage{yhmath}
\usepackage{cancel}
\usepackage{color}
\usepackage{array}
\usepackage{multirow}
\usepackage{amssymb}
\usepackage{gensymb}
\usepackage{tabularx}
\usepackage{extarrows} 
\usepackage{booktabs}
\usetikzlibrary{fadings}
\usetikzlibrary{patterns}
\usetikzlibrary{shadows.blur}
\usetikzlibrary{shapes}

\usepackage{color}
\definecolor{LinkColor}{rgb}{0.75,0.0,0.2}

\usepackage{hyperref}
\hypersetup{
	pdfauthor={good guys},
	pdftitle={good title},
	colorlinks=true,
	citecolor=LinkColor,
	linkcolor=LinkColor,
	urlcolor=LinkColor,
}

\usepackage{listings}
\definecolor{lightgray}{gray}{1}

\usepackage{theorem}

\usepackage{tikz}

\newcommand{\nc}{\newcommand}
\nc{\braoprket}[3]{\langle#1|#2|#3\rangle}
\nc{\opn}[1]{\operatorname{#1}}
\nc{\avg}[1]{\langle#1\rangle}
\nc{\ketbrasame}[1]{|#1\rangle\!\langle#1|}
\nc{\swap}{\opn{SWAP}}
\nc{\E}{\mathbb{E}}
\nc{\Var}{\opn{Var}}
\nc{\dg}{\dagger}

\usepackage[normalem]{ulem}

\begin{document}

\title{Generalized Li-Haldane Correspondence in Critical Dirac-Fermion Systems}

\author{Yuxuan Guo}
\thanks{These two authors contributed equally to this work.}
\affiliation{Department of Physics, University of Tokyo, 7-3-1 Hongo, Bunkyo-ku, Tokyo 113-0033, Japan}

\author{Sheng Yang}
\thanks{These two authors contributed equally to this work.}
\affiliation{Institute for Advanced Study in Physics and School of Physics, Zhejiang University, Hangzhou 310058, China}

\author{Xue-Jia Yu}
\email{xuejiayu@eitech.edu.cn}
\affiliation{Eastern Institute of Technology, Ningbo 315200, China}
\affiliation{Department of Physics, Fuzhou University, Fuzhou 350116, Fujian, China}
\affiliation{Fujian Key Laboratory of Quantum Information and Quantum Optics,
College of Physics and Information Engineering,
Fuzhou University, Fuzhou, Fujian 350108, China}

\begin{abstract}
Topological phenomena in quantum critical systems have recently attracted growing attention, as they go beyond the traditional paradigms of condensed matter and statistical physics. However, a general framework for identifying such nontrivial phenomena, particularly in higher-dimensional systems, remains insufficiently explored. In this work, we propose a universal fingerprint for detecting nontrivial topology in critical free-fermion systems protected by global on-site symmetries. Specifically, we analytically establish an exact relation between the bulk entanglement spectrum and the boundary energy spectrum at topological criticality in arbitrary dimensions, demonstrating that the degeneracy of edge modes can be extracted from the bulk entanglement spectrum. These findings, further supported by numerical simulations of lattice models, provide a universal fingerprint for identifying nontrivial topology in critical free-fermion systems. 

\end{abstract}

\maketitle

\section{Introduction}
Topological phases protected by global symmetries, known as symmetry-protected topological (SPT) phases, have attracted significant attention over the past two decades~\cite{Hasan2010RMP, Qi2011RMP,senthil2015symmetry, Wen2017RMP, Gu2009PRB, Chen2013PRB}. These phases cannot be characterized by local order parameters and feature robust topological edge states that hold promise for fault-tolerant quantum computing~\cite{Nayak2008RMP, AdyStern2013Science}. Traditionally, it has been widely believed that SPT physics is well-defined only in incompressible systems with a bulk energy gap~\cite{shen2012topological,bernevig2013topological} and thus is relatively well understood. However, recent advances~\cite{YU20261,Keselman2015PRB,Scaffidi2017PRX,Verresen2018PRL,Parker2018PRB,verresen2020topologyedgestatessurvive,Verresen2021PRX,Thorngren2021PRB,Duque2021PRB,Umberto2021SciPost,Yu2022PRL,Ye2022SciPost,Hidaka2022PRB,verresen2024higgscondensatessymmetryprotectedtopological,Mondal2023PRB,Wen2023PRB,Nathanan2023SciPost_a,Nathanan2023SciPost_b,thorngren2023higgscondensatessymmetryprotectedtopological,Li2023PRB,Yu2024PRL,li2024noninvertiblesymmetryenrichedquantumcritical,Su2024PRB,ando2024gauginglatticegappedgaplesstopological,Prembabu2024PRB,Li2024SciPost,wen2024topologicalholographyfermions,huang2024fermionicquantumcriticalitylens,wen2025stringcondensationtopologicalholography,Zhang2024PRA,Zhong2024PRA,Yu2024PRB,yu2025gaplesssymmetryprotectedtopologicalstates,Wen2025PRB,zhong2025quantumentanglementfermionicgapless,wen2025topologicalholography21dgapped,bhardwaj2025gaplessphases21dnoninvertible,Li2025SciPost,Huang2025SciPost,yang2025deconfinedcriticalityintrinsicallygapless,tan2025exploringnontrivialtopologyquantum,Zhou2025CP,Cardoso2025PRB,Flores2025PRL,Yang2025CP,chou2025ptsymmetryenrichednonunitarycriticality,prembabu2025multicriticalitypurelygaplessspt,deng2026anomalousdynamicalscalingtopological,yang2026topologicalquantumcriticalityquasiperiodic} have challenged this belief by revealing topological phenomena in gapless quantum critical systems, now known as topologically nontrivial quantum critical points (QCPs)~\cite{YU20261,Verresen2021PRX,Yu2022PRL} or gapless SPT (gSPT) states~\cite{Keselman2015PRB,Scaffidi2017PRX}. This extension has recently attracted considerable attention, as it reveals topological properties beyond gapped systems, including nontrivial conformal boundary conditions~\cite{Yu2022PRL, Parker2018PRB}, algebraically localized edge modes~\cite{Verresen2021PRX, Yang2025CP}, universal bulk-boundary correspondence~\cite{Yu2024PRL, Zhang2024PRA, Zhou2025CP}, and intrinsically gapless topological phases without gapped counterparts~\cite{Thorngren2021PRB,Li2025SciPost,yang2025deconfinedcriticalityintrinsicallygapless}.

Despite their importance, the identification of gSPT states has so far relied mainly on numerical simulations and has been largely limited to one-dimensional (1D) systems, owing to the absence of a unified analytical framework and efficient numerical algorithms for higher-dimensional, gapless many-body systems. gSPT physics can also emerge at QCPs between insulating phases with different topological indices~\cite{Verresen2018PRL,verresen2020topologyedgestatessurvive, Flores2025PRL}. Recent experimental advances, such as quantum anomalous Hall insulators with tunable Chern numbers, enable the study of gSPTs in topological insulator multilayers~\cite{PhysRevLett.128.216801,Deng2021} and Moiré superlattices~\cite{Polshyn2020,Saito2021}. However, unlike gapped states characterized by well-defined topological invariants, these critical states possess ill-defined invariants due to singular touching points in parameter space and therefore lack clear fingerprints for detecting nontrivial topology, especially in higher dimensions.

Nevertheless, over the past few decades, quantum entanglement has been well established as a crucial tool for characterizing gapped topological phases, from fractional quantum Hall states~\cite{Li2008PRL,Lauchli2010PRL,Regnault2009PRL,Thomale2010PRL} to both free-fermion and interacting topological systems~\cite{Fidkowski2010PRL,Pollmann2010PRB,Turner2010PRB,Turner2011PRB,Qi2012PRL,Cho_2017}. 
A remarkable advance came from Li and Haldane~\cite{Li2008PRL,Chandran2011PRB,Chandran2014PRL,Hsieh2014PRL}, who recognized that beyond entanglement entropy, more refined entanglement structures provide critical insight for understanding topological phases. 
Specifically, for a given subsystem $A$, one can define an entanglement Hamiltonian (EH) $\hat K_A$ from its reduced density matrix via $\hat{\rho}_A=\text{e}^{-\hat K_A}$.
The Li-Haldane conjecture states that there is a one-to-one correspondence between the entanglement spectrum (spectrum of $\hat{K}_{A}$) and the low-energy spectrum of the system's physical boundary.

Though initially proposed for gapped phases of matter, the conjecture has demonstrated broader applicability. Recent research suggests that the Li-Haldane conjecture may be more universal and apply to even gapless states than originally conceived. For instance, Refs.~\cite{Metlitski2011, PhysRevB.84.165134} demonstrate that information of Goldstone modes can be extracted from the entanglement spectrum. Furthermore, recent numerical investigations reveal universal structures in the entanglement spectrum at QCPs in both ladder and 2D quantum systems~\cite{Wang2025,ywk9-n2ds, Liu2025, Mao2025, Zhu2025, Li2024}. Notably, the authors and collaborators recently extended the conjecture to gSPT states in 1D interacting spin chains~\cite{Yu2024PRL}. Although formal results from algebraic quantum field theory \cite{Dalmonte2022} provide some insight into the EH, analytical results regarding the EH of gapless states in higher dimensions are particularly scarce. This motivates our effort to develop EH-based diagnostic methods for these systems in general dimensions.


This work serves to fill in that gap by proposing a universal fingerprint for identifying nontrivial boundary degeneracies in free-fermion topologically nontrivial QCPs across \emph{arbitrary dimensions}. Specifically, we analytically demonstrate an exact correspondence between the bulk entanglement spectrum and the boundary energy spectrum for critical free-fermion models in general dimensions. We show that boundary degeneracies at critical points can be extracted from the bulk entanglement spectrum, and we further verify its robustness against strong disorder and interaction. These findings, supported by lattice-model simulations, constitute a gapless generalization of the Li-Haldane bulk-boundary correspondence and, to our knowledge, the first analytical progress of such a correspondence in high-dimensional topological critical systems. Our work establishes entanglement-based diagnostics as a universal and distinctive fingerprint for identifying topological phenomena in critical free-fermion systems across arbitrary dimensions, offering a promising avenue for addressing fundamental challenges in the study of gapless topological phases of matter.

\section{Analytical Results for the Entanglement Spectrum and Boundary Modes}
A gSPT phase can be understood as a CFT wherein symmetries and anomalies preclude the existence of trivial boundary conditions at any physical boundary or interface. Similarly, evaluating the reduced density matrix requires introducing a UV cutoff as the entangling \cite{Ohmori:2014eia,fsg7-bs7q}, which is inevitably subjected to the exact same symmetry and anomaly constraints. This profound parallel strongly suggests that the Li-Haldane correspondence may remain valid even for critical systems governed by CFTs.

\begin{figure}[thbp]
  \centering
  \includegraphics[width=1.0\linewidth]{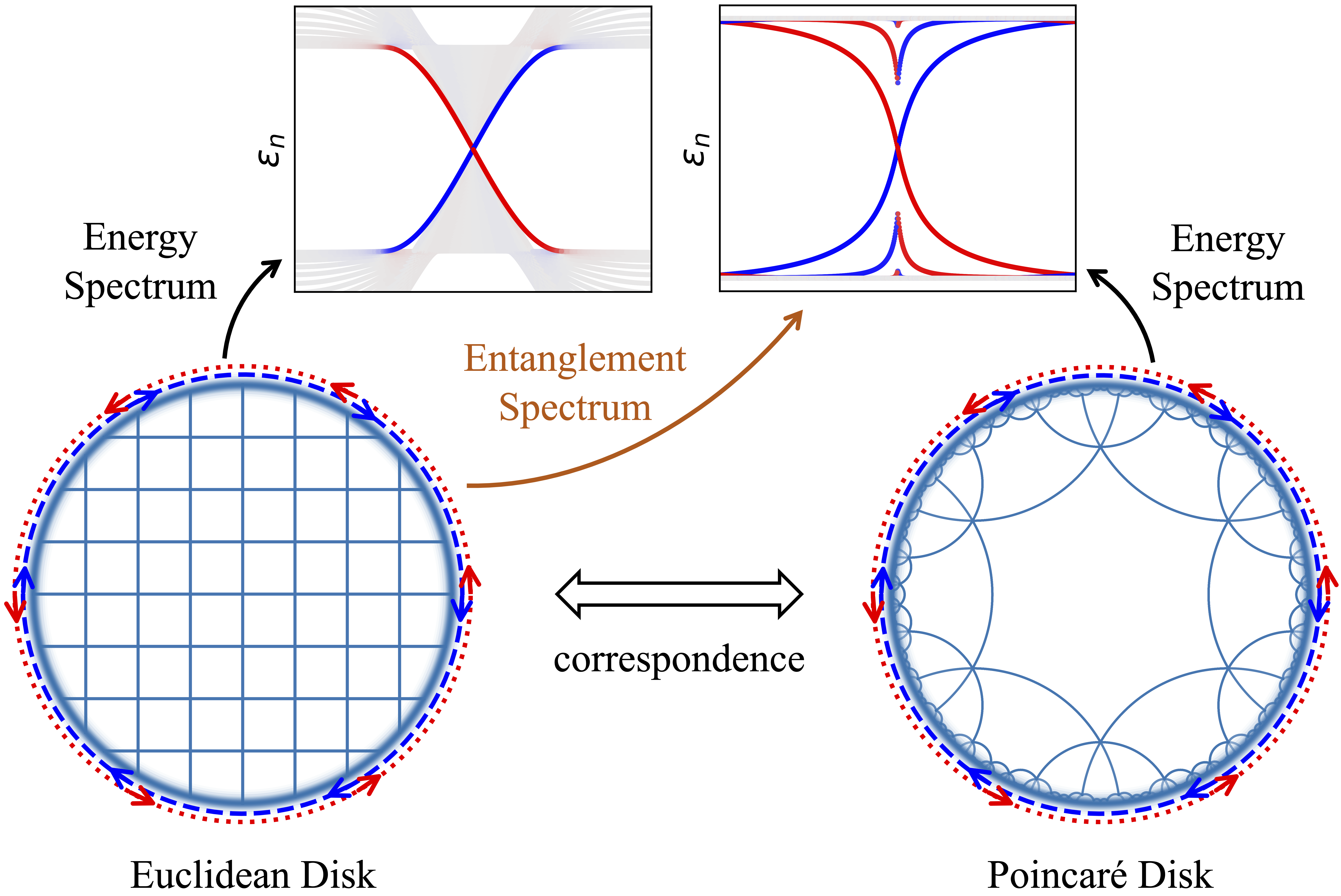}
  \caption{This figure illustrates the correspondence between the energy spectrum of a hyperbolic disk and the entanglement spectrum of a flat disk. Due to the suppression of bulk contributions in the hyperbolic geometry, the low-energy modes of the continuous bulk spectrum are no longer preserved. Since the underlying spatial metrics differ only by a Weyl transformation, the exact zero modes originating from the boundary spectrum remain robustly preserved.}
  \label{fig:poincare}
\end{figure}

To establish a concrete connection between entanglement problem and BCFT problem, we consider a spherical subregion $
A=\{\bm{x}\in\mathbb{R}^d \mid |\bm{x}|<R,\ t=0\},
$ on the $t=0$ time slice in  Euclidean spacetime with metric $ds^2_{\mathbb{R}^{d+1}}=d\tau^2+dr^2+r^2 d\Omega_{d-1}^2$, where $d\Omega_{d-1}^2$ is the metric of unit sphere, whose boundary $\partial A$ defines the entangling surface, regulated by a UV cutoff $\epsilon$. Within the replica formalism, $\Tr \hat{\rho}_A^n$ is represented by a path integral on an $n$-fold branched cover, which is conformally equivalent to $S^1_{2\pi n R}\times {\mathbb{H}}^d$ \cite{Casini2011HolographicEE}, where $\mathbb{H}^d$ is hyperbolic space in $d$ dimension. Introducing coordinates $(\tau,u,\Omega_{d-1})$, the conformal mapping can be written explicitly as
\begin{equation}
t = R\,\frac{\sinh(\tau/R)}{\cosh u+\cosh(\tau/R)},
\quad
r = R\,\frac{\sinh u}{\cosh u+\cosh(\tau/R)},
\label{eq:ball-hyperbolic-map}
\end{equation}
while leaving the angular coordinates on $S^{d-1}$ unchanged. Under this transformation, the flat metric becomes
$
ds^2=\Omega^{-2}\left(d\tau^2 + R^2\left(du^2+\sinh^2 u\,d\Omega_{d-1}^2\right)\right),
$
where $\Omega=\cosh u+\cosh(\tau/R)$ is a Weyl factor. Removing this factor yields the metric on $S^1\times \mathbb{H}^d$, we get
$
ds^2_{S^1\times \mathbb{H}^d}=d\tau^2 + R^2\left(du^2+\sinh^2 u\,d\Omega_{d-1}^2\right),
$
with a cutoff $u_{\text{max}}\sim\log\frac{2R}{\epsilon}$ imposed by the entanglement surface. Consequently, the entanglement Hamiltonian for region $A$ is unitarily equivalent to a Hamiltonian defined on $\mathbb{H}^d$ with an asymptotic boundary at $u=u_{\text{max}}$ up to an overall Casimir constant (see Fig.~\ref{fig:poincare} for an illustration)
\begin{gather}
    \hat{K}_A=2\pi R {U}^\dagger\hat{H}_{\mathbb{D}}U+\Lambda \,.
\end{gather}

Furthermore, the spatial metric
$
ds^2_{\mathbb{H}^{d}}=R^2\left(du^2+\sinh^2 u\,d\Omega_{d-1}^2\right)
$
can be mapped to the standard Poincar\'e disk via $\rho=\tanh\frac{u}{2}$, yielding
\begin{equation}
ds^2_{\mathbb{H}^{d}}=\frac{4R^2}{(1-\rho^2)^2}\left(d\rho^2+\rho^2 d\Omega_{d-1}^2\right) \,.
\end{equation}

In 1+1D, the hyperbolic disk degenerates into a line segment. The CFT energy spectrum on this segment can be exactly solved, which allows us to obtain an analytical expression for the entanglement spectrum in 1+1D. As a simple application, we consider the 1+1D case to see if we can recover the previous results in Refs. In this case, the hyperbolic space $\mathbb{H}^1$ lacks intrinsic curvature and trivially reduces to a flat 1D line segment. The spatial metric simplifies to $ds^2_{\mathbb{H}^1} = R^2 du^2$, representing a flat 1D segment with an effective length determined by the entanglement cut: $L_{\text{eff}} = \int_{-u_{\text{max}}}^{u_{\text{max}}} R \, du \simeq 2R \log(2R/\epsilon)$. For a general 1+1D CFT defined on such a finite strip with conformal boundary conditions, the exact energy spectrum is analytically known to be $E_n = \frac{\pi}{L_{\text{eff}}} \left( \Delta_n - \frac{c}{24} \right)$, where $c$ is the central charge and $\Delta_n$ are the scaling dimensions of the boundary primary operators.  Since the entanglement Hamiltonian $\hat{K}_A$ is related to the physical Hamiltonian on $\mathbb{H}^1$ evaluated over the Euclidean thermal circle of circumference $2\pi R$, the entanglement spectrum $\xi_n$ is simply given by $\xi_n = 2\pi R E_n$. Substituting $L_{\text{eff}}$, we elegantly recover the analytical result for the 1+1D entanglement spectrum :
\begin{equation}
    \xi_n = \frac{\pi^2}{\log(2R/\epsilon)}\left(\Delta_n - \frac{c}{24}\right).
\end{equation} This result is consistent with previous literature \cite{Cardy:2016fqc, yu2025gaplesssymmetryprotectedtopologicalstates}.

For higher-dimensional cases, although finding exact solutions for general CFTs on hyperbolic disk remains intractable, the situation significantly simplifies for Dirac fermions when many-body CFTs admit a single particle picture. The hyperbolic metric is Weyl-equivalent to that of the flat disk $\mathbb{D}^d$. Therefore, if the Dirac Hamiltonian on the flat disk satisfies $\slashed{\hat{D}}_{\mathbb{D}}{\psi} = E{\psi}$ and admits a zero-energy mode, the transformation properties of the Dirac operator under Weyl rescaling guarantee that the corresponding operator ${\slashed{\hat{D}}}_{\mathbb{H}}\tilde{\psi} = E\tilde{\psi}$ on $\mathbb{H}^d$ also supports a zero mode (see Appendix for details). As a result, the entanglement spectrum inherits the degeneracy arising from the zero modes of the edge spectrum. This rigorously establishes a correspondence between the boundary zero modes in the entanglement spectrum and those of a $(d+1)$-dimensional Dirac-fermion gSPT at criticality.

\section{Model and method}
At the critical point separating insulating or superconducting phases with topological indices $n$ and $n+1$ . Although the low-energy effective theories describing different transitions appear universally as massless Dirac fermions, the relevant topological degeneracy originates from the fact that, upon introducing a boundary, the symmetry-enforced massless Dirac equation admits nontrivial conformal boundary conditions. These distinct boundary conditions can support topological zero modes, thereby encoding the underlying topological structure at criticality. And this is a minimal example of gSPT in general dimension.

To realize gSPT physics in these free-fermion systems with topologically protected zero edge modes, the Hamiltonian needs to undergo a continuous phase transition between distinct topological insulating phases characterized by nonzero topological invariants~\cite{verresen2020topologyedgestatessurvive}. This requires the models to possess certain global symmetries such that the Hamiltonian falls into a symmetry class characterized by a $\mathbb{Z}$-valued topological invariant in the Altland-Zirnbauer (AZ) classification~\cite{Altland1997PRB}. We use $\hat{H}_{\alpha}$ to denote a Hamitonian with toloplogical number $\alpha\in\mathbb{Z}$.  Specifically, in 1D, we may choose $
\hat{H}^{\text{1d}}_{\alpha}(k) =
\begin{pmatrix}
\hat{c}_{A,k}^\dagger & \hat{c}_{B,k}^\dagger
\end{pmatrix}
\left[
\cos(\alpha k)\, \sigma_x + \sin(\alpha k)\, \sigma_y
\right]
\begin{pmatrix}
\hat{c}_{A,k} \\ \hat{c}_{B,k}
\end{pmatrix}
$, which belongs to the $\text{AIII}$ symmetry class. In 2D, an example is $\hat{H}^{\text{2d}}_{\alpha}(\mathbf{k}) = \begin{pmatrix}
\hat{c}_{A,\mathbf{k}}^\dagger & \hat{c}_{B,\mathbf{k}}^\dagger
\end{pmatrix}[\sin(\alpha k_x)\sigma_{x} - \sin(k_y)\sigma_y + (1-\cos(\alpha k_x) - \cos(k_y)) \sigma_z]\begin{pmatrix}
\hat{c}_{A,\mathbf{k}} \\ \hat{c}_{B,\mathbf{k}}
\end{pmatrix}$, which belongs to $\text{C}$ symmetry class.
Linear combinations of different $\hat H_{\alpha}$ can realize phase transitions between different topological insulators or superconductors~\cite{Verresen2017PRB}. For example, $ \hat H^\text{1d}_1 + \hat H^\text{1d}_2$ corresponds to the critical point that separates two gapped phases with winding numbers $\omega=1$ and $\omega=2$, featuring edge states in a nontrivial topological universality class~\cite{Verresen2017PRB, Verresen2018PRL, Verresen2021PRX, Yu2022PRL}. Similarly, $(a\hat H^\text{2d}_0+b\hat H^\text{2d}_1+c\hat H^\text{2d}_2)/(a+b+c)$ realizes a Chern insulator transition between Chern number $\mathcal C=1$ and $\mathcal C=2$ when $b=a+c$ and $c>a$, supporting chiral edge states~\cite{verresen2020topologyedgestatessurvive, Zhou2025CP}.

In the main text, we focus on 1D and 2D models as illustrative examples, while the more involved 3D case is discussed in the Appendix. We thus obtain a set of critical Hamiltonians constructed as linear combinations of $\hat H_{\alpha}$, which exhibit topologically protected edge states. Thanks to the exactly solvable nature of free-fermion models, we can easily compute the entanglement spectrum using the correlation matrix method (see Appendix B and Refs.~\cite{chung2001prb,cheong2004prb,Peschel_2009} for details), which enables the general analysis presented in the following sections.

\section{Numerical results}
\begin{figure*}[t]
  \centering
  \includegraphics[width=0.7\linewidth]{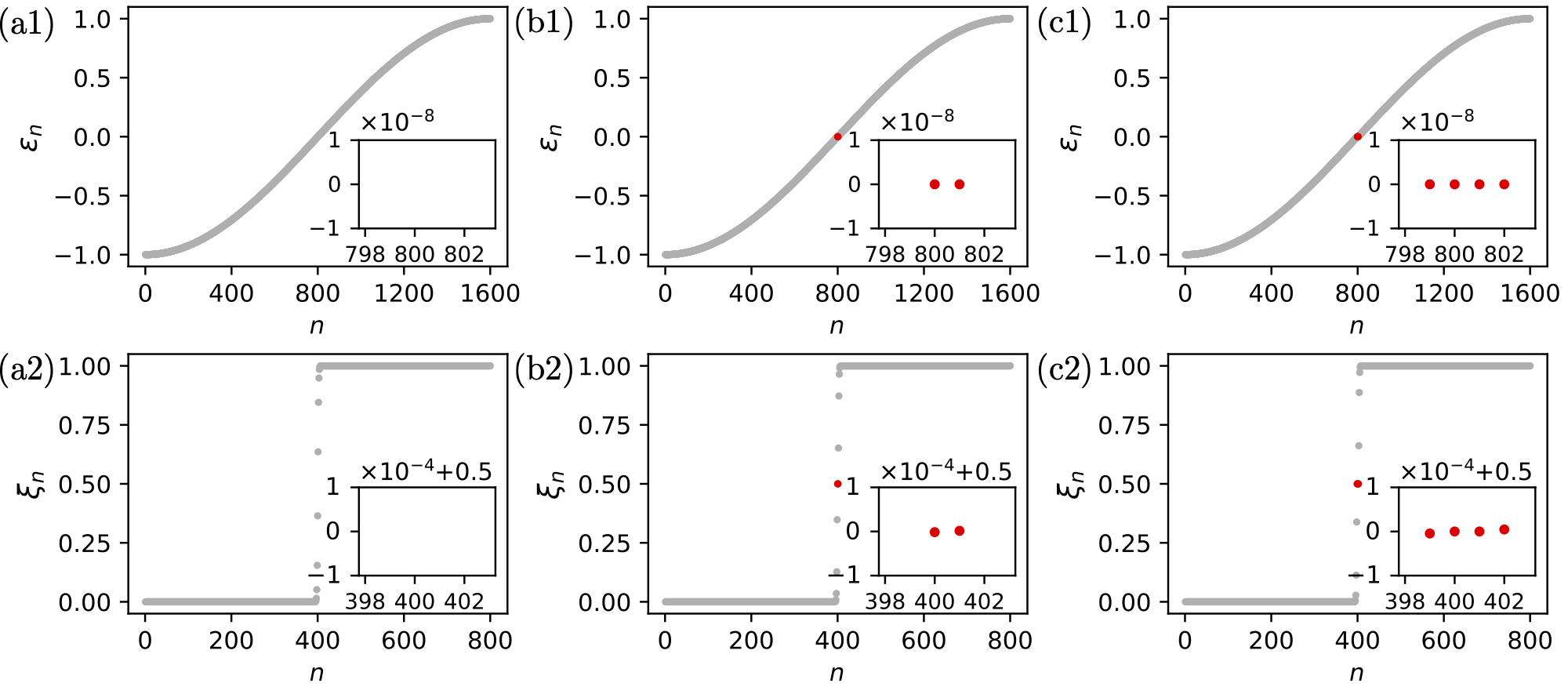}
  \caption{The boundary energy spectrum for the Hamiltonian (a1) $(\hat{H}_{0}^{\text{1d}} + \hat{H}_{1}^{\text{1d}})/2$, (b1) $(\hat{H}_{1}^{\text{1d}} + \hat{H}_{2}^{\text{1d}})/2$, and (c1) $(\hat{H}_{2}^{\text{1d}} + \hat{H}_{3}^{\text{1d}})/2$, respectively, under open boundary conditions. (a2), (b2) and (c2) are the corresponding bulk entanglement spectrum with an equal bipartition of the chain. Insets provide magnified views of specific regions to highlight the topological degenerate edge modes (red circles). The system size is $L = 800$.}
  \label{fig:1d_spectra}
\end{figure*}

To verify the analytical predictions, we now turn to lattice simulations for critical free-fermion models across different dimensions. 
In this setting, the bulk entanglement spectrum can be obtained from the gaussian state method (see Appendix B and Refs.~\cite{chung2001prb,cheong2004prb} for details)~\footnote{We either use periodic or antiperiodic boundary conditions, in our actual calculations of the bulk entanglement spectrum, to avoid possible numerical issues. For example, $(\hat{H}_{1}^{\text{1d}} + \hat{H}_{2}^{\text{1d}}) / 2$ has two exact zero-energy levels under periodic boundary conditions, which can cause ambiguity in the definition of the ground state by filling half of the lower-lying levels. To this end, we adopt the antiperiodic boundary condition to extract bulk entanglement spectra for 1D models.}. 
We first consider linear combinations of $\hat{H}_{\alpha}^{\text{1d}}$ and $\hat{H}_{\alpha+1}^{\text{1d}}$, i.e., $( \hat{H}_{\alpha}^{\text{1d}} + \hat{H}_{\alpha+1}^{\text{1d}} ) / 2$, that realize topologically distinct QCPs for different $\alpha$. 
Specifically, in 1D lattice models, the critical points are topologically trivial for $\alpha=0$ and topologically nontrivial for $\alpha=1$ and $2$, as confirmed by both the boundary energy spectra and the bulk entanglement spectra shown in Fig.~\ref{fig:1d_spectra}. 
We unambiguously demonstrate that at topologically nontrivial criticality, the boundary energy spectrum and the bulk entanglement spectrum exhibit the same boundary-state degeneracy (the number of red circles in the insets), providing numerical evidence for the correspondence between the two spectra. 
Moreover, the associated boundary degeneracies are $2$ and $4$ for $\alpha=1$ and $\alpha=2$, respectively, in exact agreement with theoretical predictions.
We note that $\xi_{n} = 1/2$ corresponds to the zero modes of the EH, therefore, we focus on the correspondence between $\epsilon_{n} = 0$ and $\xi_{n} = 1/2$ (See Appendix). Furthermore, the exact correspondence between the bulk entanglement spectrum and the boundary energy spectrum persists in the presence of symmetry-preserving disorder and interactions; see Appendix D for a detailed discussion.

\begin{figure}[htbp]
  \centering
  \includegraphics[width=1.0\linewidth]{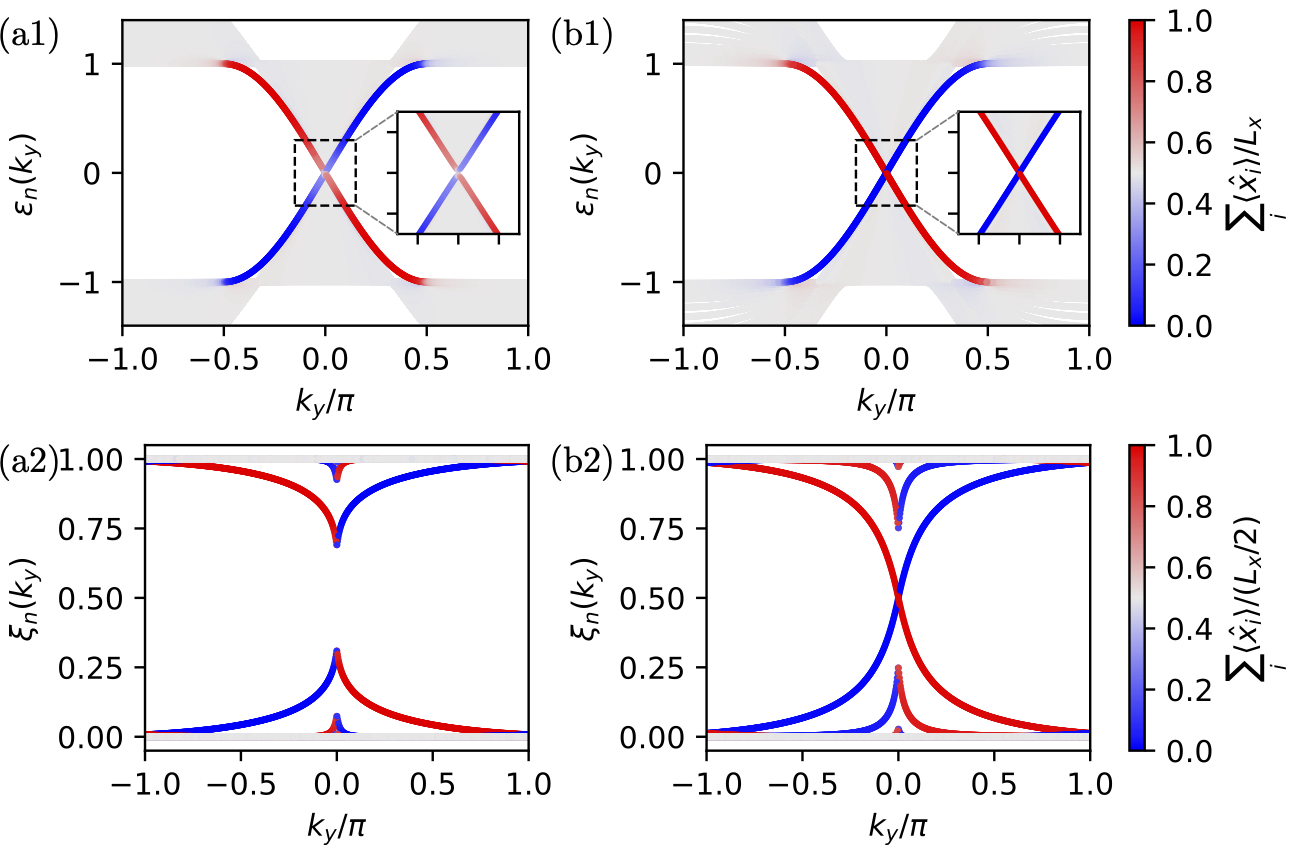}
  \caption{The energy spectrum of the critical 2D lattice model $(a\hat{H}^{\text{2d}}_{0}+b\hat{H}^{\text{2d}}_{1}+c\hat{H}^{\text{2d}}_{2})/(a+b+c)$ for (a1) $(a,b,c) = (5,6,1)$ and (b1) $(a,b,c) = (1,6,5)$, respectively. The $x$-direction is open while the $y$-direction is periodic. Insets provide zoomed-in views near the zero-energy point. In the trivial case (a1), the color of the branches fades to gray as $k_y$ approaches $0$; the faint residual red/blue hue arises from finite-size effects. (a2) and (b2) are the corresponding bipartite bulk entanglement spectrum. The entanglement cut is taken along the $x$-direction. The color coding indicates the normalized mean real-space position along the $x$-direction of each eigenstate. The blue (red) color represents the left (right) edge modes while the light gray indicates the bulk modes. The system size is $L_{x} = 40$ and $L_{y} = 800$.}
  \label{fig:2d_spectra}
\end{figure}

Beyond 1D cases, we consider topologically distinct Chern insulator transitions in 2D.
By tuning the coefficients $\{a,b,c\}$ of linear combinations of the 2D Hamiltonians $\hat{H}^\text{2d}_\alpha$ ($\alpha=0,1,2$), the model realizes Chern critical points that host chiral edge states~\cite{verresen2020topologyedgestatessurvive}. 
To reveal the associated topology in the entanglement spectrum and test its correspondence with the boundary energy spectrum, we impose open boundaries along $x$-direction and periodic boundaries along $y$-direction, and $k_y$ remains a good quantum number. Figs.~\ref{fig:2d_spectra} (a1) and (b1) show the energy spectra at representative trivial and nontrivial Chern critical points. 
For the nontrivial transition between Chern numbers $1$ and $2$ [Fig.~\ref{fig:2d_spectra} (b1)], the expected chiral edge states are clearly visible. 
In contrast, for the trivial transition between Chern numbers $0$ and $1$ [Fig.~\ref{fig:2d_spectra} (a1)], there are actually no chiral edge modes as evidenced by the color coding fading out near the gapless point $k_{y} = 0$ (corresponding to delocalized bulk states). 

Next, we compute the bulk entanglement spectrum as shown in Figs.~\ref{fig:2d_spectra} (a2) and (b2). 
Unlike the boundary energy spectrum, which does not clearly distinguish between topologically trivial and nontrivial critical points at first glance, the bulk entanglement spectrum exhibits a clear distinction: 
robust chiral edge states appear only at topologically nontrivial criticality. 
This not only provides numerical confirmation of the exact correspondence between energy and entanglement spectra in 2D, but also highlights the bulk entanglement spectrum as a more reliable fingerprint for identifying nontrivial topology in quantum critical systems. Furthermore, because the EH in our discussion is gapped and has a meaningful correspondence with the gapless physical Hamiltonian, it enables us to assign a well-defined topological invariant to critical free-fermion systems. In this sense, the bulk entanglement spectrum reveals fundamentally new information beyond boundary energy spectrum. 
In Appendix E, we extend this analysis to 3D, further confirming the correspondence between boundary energy and bulk entanglement spectra, thereby generalizing the Li-Haldane conjecture to critical free-fermion systems in arbitrary dimensions.

\section{Discussion and concluding remarks}
To summarize, we have developed a universal theory of the entanglement spectrum for identifying topological phenomena at QCPs in free-fermion systems across arbitrary dimensions. Specifically, for a broad class of topological free-fermion lattice models protected by on-site symmetries, we analytically establish an exact correspondence between the bulk entanglement spectrum and the boundary energy spectrum at criticality. More importantly, we demonstrate that the bulk entanglement spectrum captures the underlying boundary degeneracies of critical systems. Our analytical results are further supported by numerical simulations on lattice models in various dimensions. These findings establish an exact bulk-boundary correspondence in topological critical fermion systems, enabling the entanglement spectrum to serve as a universal fingerprint for detecting boundary degeneracies at criticality. As such, our framework represents a first step toward addressing the more challenging identification of topologically nontrivial QCPs in general dimensions. 

Looking ahead, it would be fascinating to explore the role of lattice symmetries in topological critical systems, where subtle distinctions in bulk-boundary correspondence may manifest in the entanglement spectrum of crystalline gapless topological phases, analogous to their gapped counterparts~\cite{Turner2010PRB}. Moreover, it is worthwhile to numerically investigate the Li-Haldane correspondence in higher-dimensional interacting gSPT phases, which are expected to host much richer topological phenomena beyond the 1D setting~\cite{Yu2024PRL,Zhang2024PRA}. On the experimental side, we note that the entanglement spectrum of band models can be measured in phononic systems~\cite{Lin2024NC}, offering a promising platform to test our theoretical predictions. In addition, the entanglement spectrum of interacting gSPT states may be efficiently accessed on state-of-the-art digital quantum platforms~\cite{zache2022entanglement,Kokail2021PRL,Joshi2023Nature,Kokail2021NatPhys} via EH learning, or on analog quantum simulators through quantum variational approaches.

\textit{Acknowledgement}: 
We thank Shao-Kai Jian and Hai-Qing Lin for collaboration on related projects, and Wen-Hao Zhong for helpful discussions.
X.-J. Yu was supported by the National Natural Science Foundation of China (Grant No.12405034) and a start-up grant from Eastern Institute of Technology, Ningbo. Y.G. is financially supported by the Global Science Graduate Course (GSGC) program at the University of Tokyo. 
The work of S.Y. is supported by China Postdoctoral Science Foundation (Certificate Number: 2024M752760).




\appendix
\onecolumngrid

 \section{Details of CFT methods}

 It establishes a rigorous correspondence between the symmetry-enforced boundary zero modes and the entanglement spectrum for general gapless interacting systems at criticality shown in Ref \cite{yu2025gaplesssymmetryprotectedtopologicalstates}.

\label{app:weyl_zero_mode}

We explicitly show that zero-energy modes of the massless Dirac Hamiltonian are invariant under spatial Weyl transformations, which is crucial for establishing the correspondence between edge modes on the flat disk $\mathbb{D}^d$ and the hyperbolic disk $\mathbb{H}^d$.

Let $g_{ij}$ and $\tilde{g}_{ij} = \Omega^2(x) g_{ij}$ be the spatial metrics on $\mathbb{D}^d$ and $\mathbb{H}^d$, respectively. Under this conformal rescaling, the massless Dirac operator transforms covariantly as:
\begin{equation}
    \slashed{\hat{D}}_H = \Omega^{-\frac{d+1}{2}} \slashed{\hat{D}}_D \Omega^{\frac{d-1}{2}}.
\end{equation}

If the flat disk Hamiltonian admits a zero-energy mode $\psi_0(x)$ satisfying $\slashed{\hat{D}}_D \psi_0 = 0$, we can define a rescaled wavefunction $\tilde{\psi}_0(x) = \Omega^{-\frac{d-1}{2}}(x) \psi_0(x)$ for the hyperbolic geometry. Applying $\slashed{\hat{D}}_H$ yields:
\begin{equation}
    \slashed{\hat{D}}_H \tilde{\psi}_0 = \left( \Omega^{-\frac{d+1}{2}} \slashed{\hat{D}}_D \Omega^{\frac{d-1}{2}} \right) \left( \Omega^{-\frac{d-1}{2}} \psi_0 \right) = \Omega^{-\frac{d+1}{2}} \slashed{\hat{D}}_D \psi_0 = 0.
\end{equation}
Thus, $\tilde{\psi}_0$ remains an exact zero-energy mode on $\mathbb{H}^d$, perfectly preserving the topological degeneracy.

In contrast, for a non-zero energy state ($\slashed{\hat{D}}_D \psi_E = E \psi_E$), the same transformation gives:
\begin{equation}
    \slashed{\hat{D}}_H \tilde{\psi}_E = \Omega^{-\frac{d+1}{2}} E \psi_E = \left( \Omega^{-1}(x) E \right) \tilde{\psi}_E.
\end{equation}
Since the Weyl factor $\Omega(x)$ is position-dependent, $\Omega^{-1}(x) E$ is not constant, meaning $\tilde{\psi}_E$ is no longer an eigenstate. This confirms that while the bulk spectrum is distorted by the geometric deformation, topologically protected zero modes remain completely immune.
\section{A brief review of Gaussian state method}

Fermionic Gaussian states play a fundamental role in many-body quantum physics, as they can be completely characterized by their two-point correlation functions~\cite{chung2001prb,cheong2004prb,Peschel_2009}. The density matrix of a fermionic Gaussian state can be expressed as:
\begin{equation}
    \hat{\rho} = \frac{1}{\mathcal{Z}} \exp\left(\frac{\mathrm{i}}{2} \sum_{i,j=1}^{2N} \Gamma_{ij} \hat{\chi}_i \hat{\chi}_j\right),
\end{equation}
where $\hat{\chi}_i$ are Majorana operators satisfying the anticommutation relation $\{\hat{\chi}_i, \hat{\chi}_j\} = 2\delta_{ij}$, and $\mathcal{Z}$ is a normalization factor. The matrix $\Gamma_{ij}$ contains information about the correlations in the state.

It is convenient to transform $\Gamma_{ij}$ into a block-diagonal form via an orthogonal transformation. This results in a collection of $2 \times 2$ blocks of the form:
\begin{equation}
    \begin{pmatrix} 0 & \lambda_k \\ -\lambda_k & 0 \end{pmatrix},
\end{equation}
where $\lambda_k \geq 0$ for $k = 1, 2, \ldots, N$. In this canonical basis, the density matrix can be rewritten as:
\begin{equation}
    \hat{\rho} = \frac{1}{\mathcal{Z}} \prod_{k=1}^N \exp\left({\mathrm{i} \lambda_k} \tilde{\hat{\chi}}_{2k-1} \tilde{\hat{\chi}}_{2k} \right),
\end{equation}
with normalization factor $\mathcal{Z} = \prod_{k=1}^N 2\cosh(\lambda_k)$. Here, $\tilde{\hat{\chi}}_i = \sum_j O_{ij} \hat{\chi}_j$ are the transformed Majorana operators, where $O$ is the orthogonal matrix that block-diagonalizes $\Gamma_{ij}$. A key advantage of the Gaussian state formalism is that it allows for straightforward computation of information-theoretic quantities. For instance, the von Neumann entropy of $\hat{\rho}$ can be expressed as:
\begin{equation}
    S_{\textrm{von}} = -\sum_{k=1}^N \left[f_k\ln f_k + (1-f_k)\ln(1-f_k)\right],
\end{equation}where $f_k = \frac{1}{1+e^{2\lambda_k}}$ represents the effective occupation number of fermions.

The covariance matrix $\Gamma_{ij}$ of a Gaussian state is determined by the two-point correlation functions of Majorana operators:
\begin{equation}
    \Gamma_{ij} = \frac{\mathrm{i}}{2}\mathrm{Tr}\big(\hat{\rho} [\hat{\chi}_i,\hat{\chi}_j]\big) = \tanh(\mathcal{H}_G)_{ij},
\end{equation}
where $[\hat{\chi}_i,\hat{\chi}_j]$ denotes the commutator of Majorana operators. To obtain the reduced density matrix of $\hat\rho$ in a subregion $A$, we only need to keep terms that are supported in subregion $A$:
\begin{gather}
    \hat\rho_A=\frac{1}{\mathcal{Z}_A}\exp{\left(\frac{\mathrm{i}}{2}\sum_{i,j \in A}(\Gamma_{A})_{ij} \hat{\chi}_i \hat{\chi}_j\right)},
\end{gather}
where $\Gamma_A$ is derived from the restriction of $\Gamma$ to region $A$, and $\mathcal{Z}_A$ is the corresponding normalization factor.

The ground state of a quadratic Hamiltonian $\hat{H} = \frac{\mathrm{i}}{2}\sum_{i,j}\mathcal{H}_{ij}\hat\chi_i\hat\chi_j$ can be regarded as a Gaussian state in the limit where $\Gamma = \tanh(\beta\mathcal{H})$ with $\beta\rightarrow\infty$. In this zero-temperature limit, the covariance matrix simplifies to:
\begin{equation}
\Gamma = -\mathrm{i}U^\dagger\, \operatorname{sgn}(D)\, U,
\end{equation}
where we denote this $\Gamma$ as the entanglement Hamiltonian $\mathcal{K}_E$. Here, $U$ is the unitary matrix that diagonalizes $\mathrm{i}\mathcal{H}$ as $U(\mathrm{i}\mathcal{H})U^\dagger = D$, and $\operatorname{sgn}(D)_{ii} = 1$ if $D_{ii} \leq 0$, and $-1$ otherwise.

Upon further diagonalization of this entanglement Hamiltonian, we obtain the reduced density matrix in the following form:
\begin{equation}
    \hat{\rho}_A = \prod_{k=1}^m\left(\frac{1}{2}+\frac{\mathrm{i}}{2}\tanh{\lambda_k}\tilde{\hat{\chi}}_{A,2k-1} \tilde{\hat{\chi}}_{A,2k}\right),
\end{equation}
where $\lambda_k$ are the eigenvalues of the entanglement Hamiltonian $\mathcal{K}_{A,E}$. In numerical calculations, we can conveniently obtain the quantities $\frac{1}{2}\pm\frac{\mathrm{1}}{2}\tanh{\lambda_k}$, which we refer to as the single-particle entanglement spectrum. The equivalence between this spectrum and the boundary energy spectrum will be demonstrated in the following subsection for free-fermion critical states.

\section{Brief Review of Low-Dimensional Topological Critical Free Fermion Models}
In one dimension (1D), the gapped phases of the A\uppercase\expandafter{\romannumeral3} class are $\mathbb{Z}$-classified, and a topological state with winding number $\alpha=n$ is described by the Hamiltonian~\cite{Verresen2018PRL,verresen2020topologyedgestatessurvive,Verresen2021PRX,Yu2022PRL}:
\begin{gather}
    \hat{H}^{\text{1d}}_{n}=\sum_i\hat{c}^\dagger_{A,i}\hat{c}_{B,i+n}+\text{h.c.}
\end{gather}

In momentum space, this Hamiltonian reads:
\begin{gather}
   \hat{H}^{\text{1d}}_{n}=\sum_{k}\hat{\Psi}_k^{\dagger}[\cos(nk)\sigma_x+\sin(nk)\sigma_y]\hat{\Psi}_k
\end{gather}
where we introduce the basis $
\hat{\Psi}_k^{\dagger} = (\hat{c}_{A,k}^\dagger,\hat{c}_{B,k}^\dagger)
$.

\begin{figure*}[htbp]
  \centering
  \includegraphics[width=1.0\linewidth]{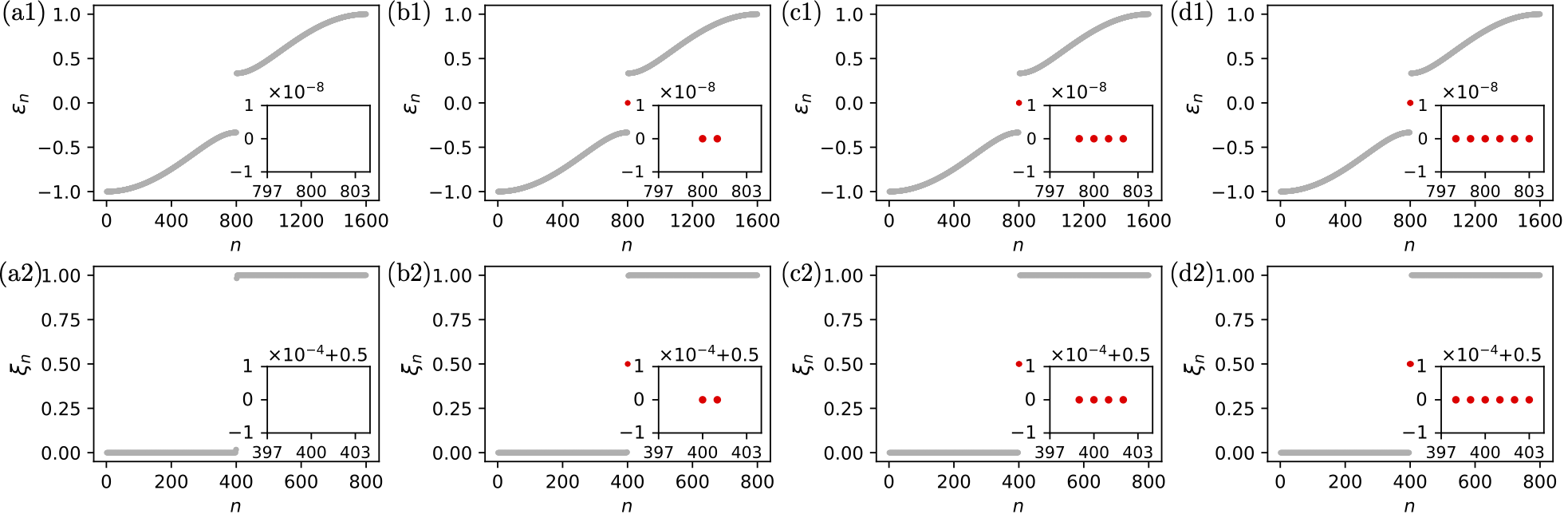}
  \caption{The boundary energy spectrum for the Hamiltonian (a1) $(2\hat{H}_{0}^{\text{1d}} + \hat{H}_{1}^{\text{1d}})/3$, (b1) $(2\hat{H}_{1}^{\text{1d}} + \hat{H}_{2}^{\text{1d}})/3$, (c1) $(2\hat{H}_{2}^{\text{1d}} + \hat{H}_{3}^{\text{1d}})/3$, and (d1) $(\hat{H}_{2}^{\text{1d}} + 2\hat{H}_{3}^{\text{1d}})/3$, respectively. (a2), (b2), (c2), and (d2) are the corresponding half-chain bulk entanglement spectrum. Insets provide magnified views of specific regions to highlight the topological degenerate edge modes (red circles). The system size is $L = 800$.}
  \label{fig:1d_gapped}
\end{figure*}

It is clear that the winding number of this Hamiltonian is $n$. A linear combination $\sum_{\alpha}c_\alpha \hat{H}^{\text{1d}}_{\alpha}$ becomes critical under two conditions. The first condition occurs when $\sum_\alpha c_\alpha = 0$, producing a gapless point at $k=0$. The second condition arises when $\sum_\alpha(-1)^\alpha c_\alpha = 0$, resulting in a gapless point at $k=\pi$. In our main text, we focus on two specific models: $(\hat{H}^{\text{1d}}_{1} + \hat{H}^{\text{1d}}_{2})/2$ and $(\hat{H}^{\text{1d}}_{2} + \hat{H}^{\text{1d}}_{3})/2$, both of which exhibit the gapless topological characteristics described above.
Fig.~\ref{fig:1d_gapped} shows the energy levels under open boundary conditions and the bulk entanglement spectrum for representative parameters within the gapped phases of the winding number $\nu = 0$ (a1-a2), $\nu = 1$ (b1-b2), $\nu = 2$ (c1-c2), and $\nu = 3$ (d1-d2), respectively.
It is clear that the topological zero edge modes (red circles in the plotting) can be fully reflected in both energy and entanglement spectra.

\begin{figure}[t]
  \centering
  \includegraphics[width=0.75\linewidth]{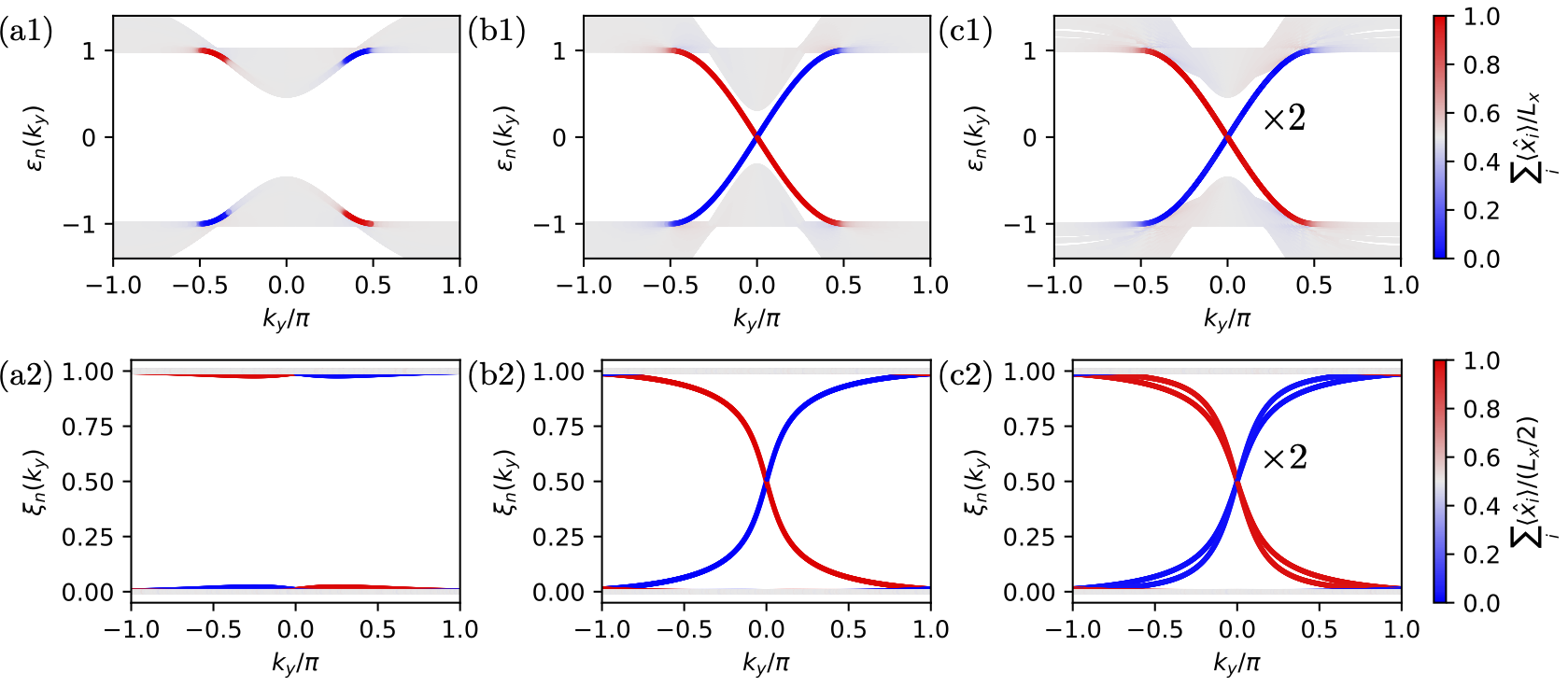}
  \caption{The energy spectrum of the gapped 2D lattice model $(a\hat{H}^{\text{2d}}_{0}+b\hat{H}^{\text{2d}}_{1}+c\hat{H}^{\text{2d}}_{2})/(a+b+c)$ for (a1) $(a,b,c) = (4,1,1)$, (b1) $(a,b,c) = (1,4,1)$, (c1) $(a,b,c) = (1,1,4)$, respectively. The $x$-direction is open while the $y$-direction is periodic. (a2), (b2), and (c2) are the corresponding bipartite bulk entanglement spectrum. The entanglement cut is taken along the $x$-direction. The color coding indicates the normalized mean position along the $x$-direction. The blue (red) color represents the left (right) edge modes while the light gray indicates the bulk modes. The system size is $L_{x} = 40$ and $L_{y} = 800$.}
  \label{fig:2d_gapped}
\end{figure}

In 2D, class A is $\mathbb{Z}$ classified Chern insulator, the lattice model corresponding to $\hat{H}^{\text{2d}}_{1}(\mathbf{k}) = \begin{pmatrix}
\hat{c}_{A,\mathbf{k}}^\dagger & \hat{c}_{B,\mathbf{k}}^\dagger
\end{pmatrix}[\sin(k_x)\sigma_{x} - \sin(k_y)\sigma_y + (1-\cos(k_x) - \cos(k_y)) \sigma_z]\begin{pmatrix}
\hat{c}_{A,\mathbf{k}} \\ \hat{c}_{B,\mathbf{k}}
\end{pmatrix}$ reads as
\begin{gather}
     \hat{H}^{\text{2d}}_{1}=\frac{1}{2} \sum_{i,j}\Big[-\text{i}\hat{c}_{A,{i+1,j}}^{\dagger}\hat{c}_{B,{i,j}}-\text{i}\hat{c}_{B,{i+1,j}}^{\dagger}\hat{c}_{A,{i,j}}+ \hat{c}_{A,{i,j+1}}^{\dagger}\hat{c}_{B,{i,j}}-\hat{c}_{B,{i,j+1}}^{\dagger}\hat{c}_{A,{i,j}} \nonumber\\
     +\hat{c}_{A,{i,j}}^{\dagger}\hat{c}_{A,{i,j}}-\hat{c}_{B,{i,j}}^{\dagger}\hat{c}_{B,{i,j}}
     -\hat{c}_{A,{i+1,j}}^{\dagger}\hat{c}_{A,{i,j}}+\hat{c}_{B,{i+1,j}}^{\dagger}\hat{c}_{B,{i,j}} \nonumber\\
     -\hat{c}_{A,{i,j}}^{\dagger}\hat{c}_{A,{i,j+1}}+\hat{c}_{B,{i,j}}^{\dagger}\hat{c}_{B,{i,j+1}}
     \Big]+\text{h.c.}
\end{gather}
The Chern number can be calculated using the formula:
\begin{equation}
\nu = \int_{\text{BZ}} \frac{d^2k}{4\pi^2}\varepsilon^{\mu\nu} \text{Tr}[(h^{-1}\partial_{\mu}h)(h^{-1}\partial_{\nu}h)],
\end{equation}
where $h$ is the flattened Hamiltonian of $H$. Notably, a topological insulator with higher Chern number $n$ can be readily constructed by replacing $i+1$ with $i+n$ in the original Hamiltonian. When we consider a linear combination $\hat{H}^{\text{2d}}=\sum_{\alpha}c_{\alpha}\hat{H}^{\text{2d}}_\alpha$, the system becomes gapless at $\mathbf{k}=(\pi,0)$ when the condition $\sum_{\alpha}(-1)^{\alpha}c_\alpha=0$ is satisfied. For the specific case where $\alpha \in \{0,1,2\}$, the Hamiltonian $(a\hat{H}^{\text{2d}}_0+b\hat{H}^{\text{2d}}_1+c\hat{H}^{\text{2d}}_2)/(a+b+c)$ describes a topological phase transition between Chern insulators with $\mathcal{C}=1$ and $\mathcal{C}=2$ when $b=a+c$ and $c>a$, supporting chiral edge states. Conversely, when $a<c$, the system realizes a trivial critical point.
Fig.~\ref{fig:2d_gapped} shows the energy levels and the bulk entanglement spectrum for representative parameters within the gapped phases of $\mathcal{C} = 0$ (a1-a2), $\mathcal{C} = 1$ (b1-b2), and $\mathcal{C} = 2$ (c1-c2), respectively.
To distinguish between the delocalized bulk states and the localized left/right edge states, we evaluate the normalized mean position along the open $x$-direction for each eigenmode. 
For a given energy eigenstate $|\psi_{n,k_{y}}\rangle$ corresponding to the energy level $\epsilon_{n,k_{y}}$, the mean position $\bar{x}_{n}(k_{y})$ of this state can be defined as
\begin{equation}
    \bar{x}_{n}(k_{y}) \equiv \frac{1}{L_{x}} \sum_{i=1}^{L_{x}} \langle \hat{x}_{i} \rangle = \frac{1}{L_{x}} \sum_{i=1}^{L_{x}} \sum_{\alpha = A,B} i |\psi_{n,k_{y}}^{\alpha}(i)|^{2} \,,
\end{equation}
where $i$ denotes the index of the unit cells along the $x$-direction, and $\psi_{n,k_{y}}^{\alpha}(i)$ represents the amplitude of the fermion on sublattice $\alpha\in\{A,B\}$ at unit cell $i$. 
According to this definition, a mean-position value close to $0$ (or $1$) corresponds to a state localized on the left (or right) boundary, whereas a value near $1/2$ indicates a bulk state. 
This methodology is also applied in the same manner to the bulk entanglement spectrum. 
In that case, we calculate the normalized mean position using the eigenstates of the entanglement Hamiltonian instead of the physical Hamiltonian.
As shown in Fig.~\ref{fig:2d_gapped}, for $\mathcal{C} = 1$ and $2$, we can easily see the existence of topological zero edge modes through both energy and entanglement spectra.

\begin{figure}[t]
  \centering
  \includegraphics[width=0.4\linewidth]{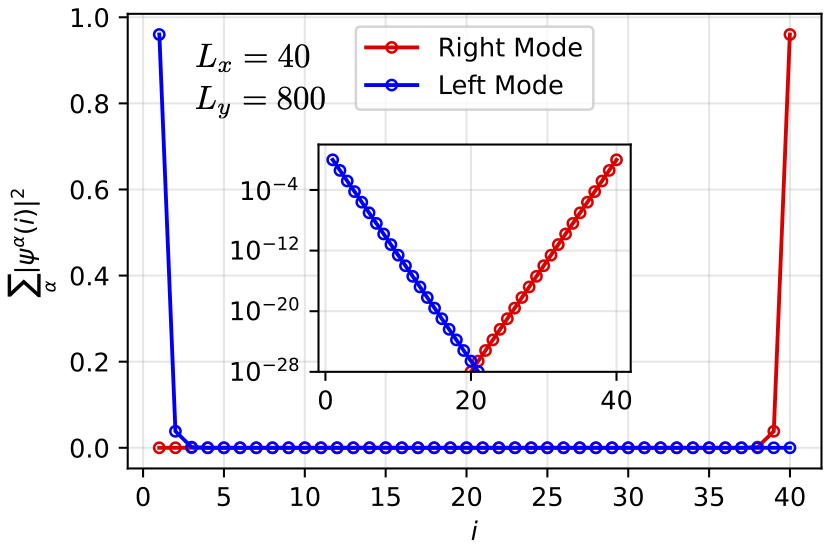}
  \caption{The spatial dependence of the density $\sum_{\alpha=A,B} |\psi^{\alpha}(i)|^{2}$ on the index $i$ of the unit cells calculated for the left and right edge modes at the momentum point $k_y$ closest to the zero-energy point of the gapless 2D lattice model $(a\hat{H}^{\text{2d}}_{0}+b\hat{H}^{\text{2d}}_{1}+c\hat{H}^{\text{2d}}_{2})/(a+b+c)$ with $(a,b,c) = (1,6,5)$. The $x$-direction is open while the $y$-direction is periodic. The semi-logarithmic plot in the inset indicates that both edge modes are exponentially localized on the boundary. The system size is $L_{x} = 40$ and $L_{y} = 800$.}
  \label{fig:2d_profile}
\end{figure}

In the main text, we have examined the energy spectrum and the corresponding bulk entanglement spectrum for gapless systems, specifically for the parameter sets $(a, b, c) = (5, 6, 1)$ and $(1, 6, 5)$. Although the normalized mean position indicates the presence of robust edge modes at the zero-energy point for the latter case, it does not explicitly reveal the nature of their localization. To address this crucial point, we calculate the spatial dependence of the density $\sum_{\alpha=A,B} |\psi^{\alpha}(i)|^{2}$ on the index $i$ of the unit cells. As shown in Fig.~\ref{fig:2d_profile}, the results demonstrate that the edge modes are perfectly exponentially localized on the boundaries.

Since the 2D Chern insulator belongs to class A (with $U(1)$ symmetry as a physical symmetry), its $\mathbb{Z}$ classification remains stable in the presence of interactions. 
We therefore introduce an interacting model that is pinned to the critical state when the interaction strength is increased, given by
\[
\hat{H}(\lambda) = \lambda \hat{H}^\text{2d}_1 + (1 - \lambda) \hat{H}^\text{2d}_2 + \hat{H}_{\text{int}},
\]
where $\hat{H}_1^\text{2d}$ and $\hat{H}_2^\text{2d}$ are dual to each other under the transformation 
$\hat{c}_{A,i,j} \leftrightarrow \hat{c}_{A,-i+3,j}$ and 
$\hat{c}_{B,i,j} \leftrightarrow \hat{c}_{B,-i,j}$. 
The interaction term invariant under this dual transformation reads
\begin{align}
\hat{H}_{\text{int}}
&= U \sum_{i,j} 
\big[
  (\hat{c}_{A,i,j}^{\dagger} \hat{c}_{B,i,j} + \hat{c}_{B,i,j}^{\dagger} \hat{c}_{A,i,j}) 
  ( \hat{n}_{A,i,j} + \hat{n}_{B,i,j} ) \notag \\
&\quad + 
  (\hat{c}_{A,i,j}^{\dagger} \hat{c}_{B,i+3,j} + \hat{c}_{B,i+3,j}^{\dagger} \hat{c}_{A,i,j}) 
  ( \hat{n}_{A,i,j} + \hat{n}_{B,i+3,j} )
\big] .
\end{align}
Under the duality transformation, the Hamiltonian satisfies 
$\hat{H}(\lambda) \leftrightarrow \hat{H}(1 - \lambda)$. 
This duality ensures that if a phase transition occurs upon varying $\lambda$, 
it must be pinned at $\lambda = 1/2$.

We also remark that although all gSPT states have gapless bulk, the energy splitting of edge degeneracies is more subtle and can exhibit either exponential or algebraic decay depending on whether the bulk hosts additional gapped degrees of freedom~\cite{Verresen2021PRX,Li2025SciPost}. Specifically, if the gapless system contains additional gapped degrees of freedom—for example, a topological quantum critical system described by a conformal field theory (CFT) stacked with a gapped SPT state~\cite{Scaffidi2017PRX}—there are no perturbations within the low-energy CFT that can couple the two degenerate ground states. Any effective interaction must therefore be mediated through the gapped degrees of freedom, leading at most to a finite-size splitting of order 
$e^{-\text{const} \times L}$~\cite{Verresen2021PRX}, 
as in our case in the main text. Conversely, if the gSPT has no gapped degrees of freedom—meaning that all symmetry sectors involved in the gapped SPT act faithfully on the low-energy gapless theory—this corresponds to a purely gapless SPT phase~\cite{Verresen2021PRX,verresen2020topologyedgestatessurvive}. In this case, the finite-size splitting is algebraic. These algebraically localized edge modes constitute one of the important aspects of gSPT physics, and to the best of our knowledge, only a few lattice realizations of purely gSPT phases currently exist in the literature~\cite{Verresen2021PRX,Yu2022PRL,Prembabu2024PRB,prembabu2025multicriticalitypurelygaplessspt}.

Moreover, Ref.~\cite{verresen2020topologyedgestatessurvive} clearly demonstrates that topological edge modes in critical free-fermion systems are always exponentially localized. Below we provide a brief argument supporting this statement~\cite{verresen2020topologyedgestatessurvive}.

The key issue is whether exponentially localized edge modes remain stable when coupled to a critical bulk and, in particular, whether such coupling can generate algebraic contributions to the edge modes in free-fermion systems. To address this, we consider the BDI symmetry class as an illustrative example and start from the effective free-fermion Majorana theory with boundary condition $H = i\int_{0}^{\infty} \tilde{\chi}(x)\partial_x \chi(x) dx, \tilde{\chi}(0)=0$, together with an exponentially localized edge mode $\chi_{\text{loc}} = \int_{0}^{\infty} e^{-\mathrm{const} x}\chi(x) dx$. Here, $\chi$ and $\tilde{\chi}$ denote the two species of Majorana operators, and $\chi_{\text{loc}}$ represents the zero-energy localized edge mode. Then we introduce some perturbations to discuss the splitting of edge modes. The dominant couplings, $i\chi_{\text{loc}}\chi(0)$ and $i\chi_{\text{loc}}\tilde{\chi}(0)$, are either forbidden by symmetry or vanish due to the imposed boundary condition. One may further consider higher-derivative irrelevant perturbations such as $i\chi_{\text{loc}}\partial_x \tilde{\chi}(0)$. However, these terms can be removed by an appropriate unitary transformation. In fact, in the non-interacting (quadratic) setting, one can show that all such perturbations can be rotated away, implying that free-fermion perturbations cannot induce algebraic decay for the edge mode. This explains why we focus exclusively on exponentially localized edge modes throughout the paper.

And regarding the effect of interactions on critical free fermion systems, as shown in Fig.~3 (c) of the main text, there exists a homomorphism from free fermion gapped SPTs, which are classified by the $K$ group, to the cobordism classification of interacting fermionic gapped SPTs (for example, in our case, the $\mathbb{Z}$ classification is reduced to $\mathbb{Z}_4$ and the homomorphism is given by $\mathbb{Z}/4\mathbb{Z}$
). This implies that there exists a finite-depth quantum circuit connecting the free fermion gapped topological phase to the corresponding interacting gapped topological phase. Therefore, no phase transition occurs when interactions are introduced properly. As interactions are turned on, a phase boundary (a critical line) exists between two distinct fermionic gapped topological phases. If we assume that the universality class of phase transition between two phases is unique, there will be no additional phase transition occurs at this phase boundary—which is indeed the case align with over numeric—we claim that the low-energy behavior of the boundary spectrum and entanglement spectrum in the critical free fermion case has a one-to-one correspondence with its interacting counterpart, and thus supports exponentially localized edge modes.

\section{Effects of disorder and interactions}
\label{sm1.5}
\begin{figure}[t]
  \centering
  \includegraphics[width=0.6\linewidth]{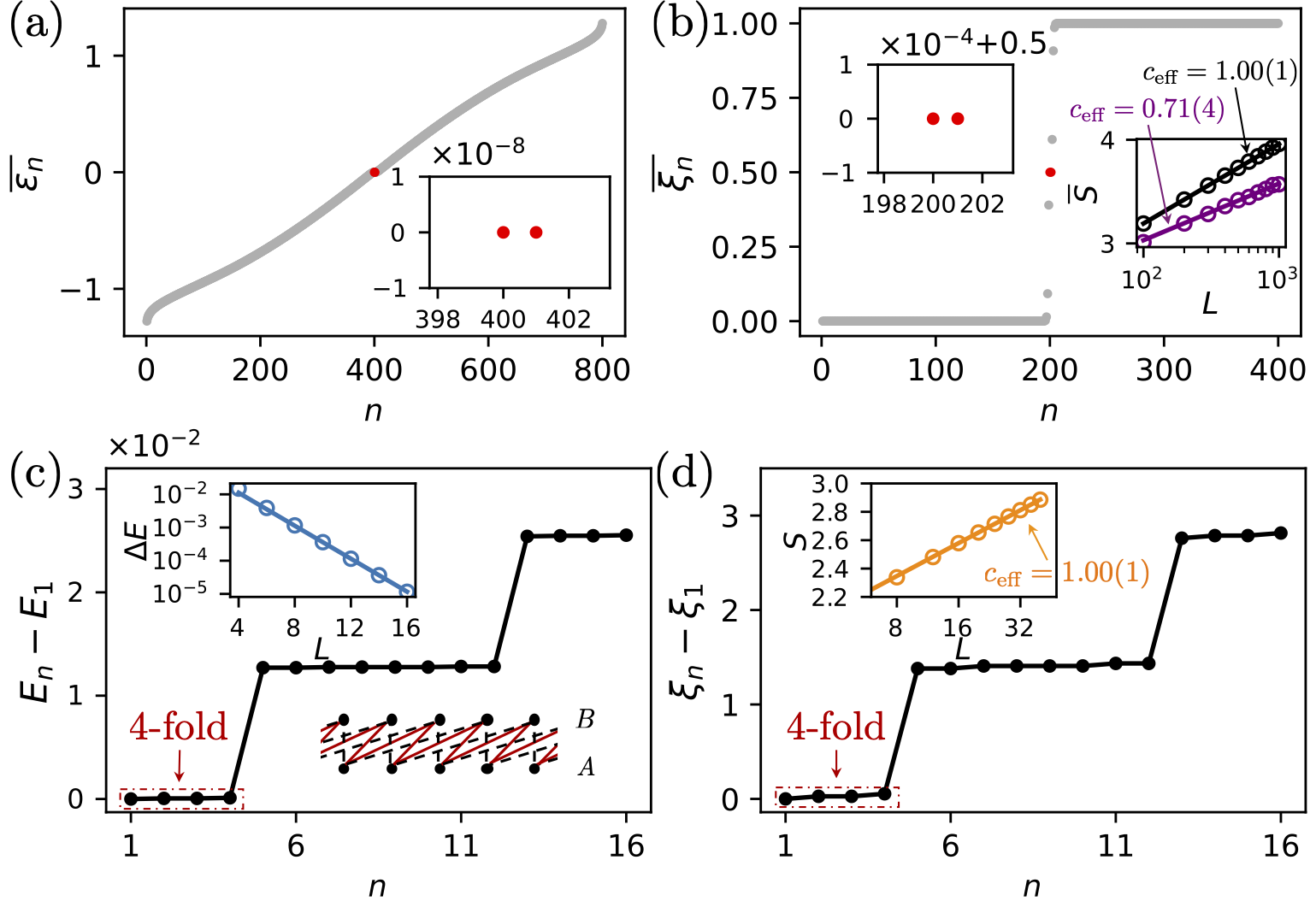}
  \caption{(a) The averaged energy spectrum $\overline{\epsilon_{n}}$ under open boundary conditions and (b) the corresponding bulk entanglement spectrum $\overline{\xi_{n}}$ of the critical 1D model at strong disorder $\delta = 0.5$\,. Insets provide magnified views of specific regions to highlight the topological degenerate edge modes (red circles) and logarithmic scalings of the half-chain entanglement entropy that suggest $c_{\text{eff}} = 1.00(1)$ for the clean case and $c_{\text{eff}} \approx \ln{2}$ for the disordered case, respectively. Results are averaged over $10^{4}$ disorder realizations for $L = 400$. (c) The many-body energy spectrum under open boundary conditions and (d) the corresponding bulk entanglement spectrum of the interacting 1D model for $L = 12$ with $U = 0.15$\,. The total number of fermions is not fixed in the calculation of the energy spectrum. The inset in (c) gives an schematic for the model, where red solid lines represent the free part of the model and the black dashed lines are interaction terms, as well as the energy splitting within the degenerate edge modes due to the interaction is exponentially small, $\Delta E \equiv E_{4} - E_{1} \sim \exp(-L/l)$ with $l \approx 1.75$\,. To resolve this finite-size energy splitting, relatively small system sizes are chosen in our calculations of the spectra. The entanglement scaling in the inset of (d) indicates that the model is critical and belongs to the same universality class of the free model. The bond dimension of the matrix product state used in the simulation is $\chi = 512$ for (c) and $\chi = 1024$ for (d).}
  \label{fig:1d_add}
\end{figure}

Topological physics in gapless systems has often been referred to as topological semimetals in the literature~\cite{Armitage2018RMP,yan2017topological}. 
Here we emphasize the crucial distinction between free-fermion topologically nontrivial QCPs and topological semimetals. 
For example, a 3D Weyl semimetal is gapless only at isolated points in momentum space, while any 2D slice that avoids these points is a gapped lower-dimensional system with nontrivial Chern numbers, giving rise to the well-known Fermi arcs that can be regarded as ``topological edge states'' in this context. 
Thus, although topological semimetals are gapless at discrete points, their ``topological edge states'' are inherent from the gapped slices, in sharp contrast to our case, where the edge states occur exactly at the gapless critical points. 
Moreover, topological semimetals rely on a \emph{momentum-dependent} energy gap and therefore would be destabilized by disorder, which mixes momentum between different gapless points~\cite{PIXLEY2021168455}. 
In contrast, the boundary degeneracies of free-fermion topologically nontrivial QCPs are robust against symmetry-preserving disorders, and importantly, the Li-Haldane correspondence remains valid, as confirmed numerically in a 1D disordered critical model as shown in Figs.~\ref{fig:1d_add} (a) and (b). 
The disorder is added directly to the real-space version of the critical Hamiltonian $(\hat{H}_{1}^{\text{1d}} + \hat{H}_{2}^{\text{1d}}) / 2$ after the Fourier transformation $\hat{c}_{A(B),k} = 1/\sqrt{L} \sum_{j} \text{e}^{\text{i} j k } \hat{c}_{A(B),j}$, and the disordered Hamiltonian is $\hat{H}_{\text{disorder}}^{\text{1d}} = \sum_{\alpha=1,2} \sum_{i} t_{i}^{(\alpha)} (\hat{c}_{A,i}^{\dagger} \hat{c}_{B,i+\alpha}^{} + \hat{c}_{B,i+\alpha}^{\dagger} \hat{c}_{A,i}^{})$, where $t_{i}^{(\alpha)}$ are sampled independently from the uniform distribution $[1-\delta, 1+\delta]$ for each $i$ and $\alpha$. 
$\delta$ is the strength of the disorder which is set to $0.5$ here. 
We have also confirmed that the critical system flows to the infinite randomness fixed point at strong disorder with effective central charge $c_{\text{eff}} = \ln{2}$; this is the so-called random singlet fixed point where the disorder flows to infinity~\cite{PhysRevLett.93.260602,PhysRevB.50.3799,PhysRevB.51.6411}. Though the disorder flows to infinity, both the boundary degeneracies of the criticality and the correspondence between boundary energy levels and bulk entanglement spectrum still remain, since our proof is purely based on real space without assuming any spacetime symmetry, and these results naturally generalize to higher-dimensional topological critical fermion systems with symmetry-preserving disorder. 

In the presence of interactions, the classification of free-fermion topological phases can be reduced, but some free-fermion critical states with nontrivial topology remain nontrivial even after interactions are introduced. This reduction implies that the free-fermion topological phase and its corresponding interacting phase belong to the same phase, and there exists a finite-depth quantum circuit connecting them. As interactions are turned on, a phase boundary exists between two distinct topological fermion phases. If we assume the universality of phase transition occurs at this phase boundary between two phases is unique, the low-energy behavior of the entanglement spectrum of gSPT in the free-fermion case has a one-to-one correspondence with its interacting counterpart. A famous example of the interaction reduction is free fermion protected by $\mathbb{Z}_2^{T}\times\mathbb{Z}_2^F$ with $\hat{T}^2=1$ in 1D (class BD\uppercase\expandafter{\romannumeral1} for free fermion), where the classification is reduced to $\mathbb{Z}_8$ due to interaction \cite{PhysRevB.81.134509,PhysRevB.83.075103,sym14071475}. With additional $U(1)$ symmetry (class A\uppercase\expandafter{\romannumeral3} for free fermion~\footnote{A SSH model can be constructed from $\alpha=2$ chain model $\hat H_{SSH}=\text{i}\sum{\hat \gamma}_n\hat{\tilde{\gamma}}_{n+2}$  and identifying $\hat c_{A,n}=\hat{{\gamma}}_{2n}+\text{i}{\hat \gamma}_{2n-1}$ and $\hat c_{B,n}=\hat {\tilde{\gamma}}_{2n-1}-\text{i}\hat {\tilde{\gamma}}_{2n}$ and the original time reversal symmetry $\hat T\hat \gamma \hat T=\hat \gamma$ and $\hat T\hat{\tilde\gamma} \hat T=-\hat{\tilde\gamma}$ acting like a sublattice symmetry on complex fermions, where $\hat T\hat c_A\hat T=\hat c_A^\dagger$ and $\hat T\hat c_B\hat T=-\hat c_B^\dagger$.}), classification is reduced to $\mathbb{Z}_4$ \cite{PhysRevLett.109.096403,PhysRevB.90.115141}. This suggests that fermionic gSPT should survive even in the presence of strong interaction. Here, we consider an interacting topological critical fermion chain $\hat{H}_{\text{I}}^{\text{1d}} = (\hat{H}_{1}^{\text{1d}} + \hat{H}_{2}^{\text{1d}}) / 2 + U \sum_{i} \big[ (\hat{c}_{A,i}^{\dagger} \hat{c}_{B,i}^{} + \hat{c}_{B,i}^{\dagger} \hat{c}_{A,i}^{}) ( \hat{n}_{A,i} + \hat{n}_{B,i} ) + (\hat{c}_{A,i}^{\dagger} \hat{c}_{B,i+3}^{} + \hat{c}_{B,i+3}^{\dagger} \hat{c}_{A,i}^{}) ( \hat{n}_{A,i} + \hat{n}_{B,i+3} ) \big]$, where we assume the particle-hole symmetry holds approximately for half filling.  
The interacting model is critical for any $U \neq 0$ due to the self-duality property~\footnote{As shown in the inset of Fig.~7 (c), the non-interacting model is analogous to the well-known Su-Schrieffer-Heeger model after suitable numbering of the sites. The dual transformation is just a single-site translation of the system along the red solid line, under which $\hat{H}_{1}^{\text{1d}} \leftrightarrow \hat{H}_{2}^{\text{1d}}$ and the chiral-symmetry-preserving interaction term remains unchanged. Here, we do not consider the interaction $ (\hat{c}_{A,i}^{\dagger} \hat{c}_{B,i+1}^{} + \hat{c}_{B,i+1}^{\dagger} \hat{c}_{A,i}^{}) ( \hat{n}_{A,i} + \hat{n}_{B,i+1} )$, which is equivalent to $ (\hat{c}_{A,i}^{\dagger} \hat{c}_{B,i+2}^{} + \hat{c}_{B,i+2}^{\dagger} \hat{c}_{A,i}^{}) ( \hat{n}_{A,i} + \hat{n}_{B,i+2} )$ after dual transformation, since it does not touch degrees of freedom of the danging fermions at the edges.}. 
Using the density matrix renormalization group method~\cite{white1992prl,white1993prb,ulrich2011ap,itensor}, we numerically compute the many-body energy spectrum and the corresponding bulk entanglement spectrum for $U=0.15$, as shown in Figs.~\ref{fig:1d_add} (c) and (d). 
It is noted that the two-fold degenerate edge modes in the single-particle spectrum indicates that the ground-state manifold of the many-body spectrum should be four-fold degenerate as we have not fixed the particle number in our calculations in Fig.~\ref{fig:1d_add} (c).
We confirm that the associated boundary degeneracies remain encoded in the bulk entanglement spectrum even when symmetry-preserving interactions of suitable magnitudes are included.
Similar behavior is also observed in 1D quantum spin systems~\cite{Yu2024PRL}. 
For higher-dimensional interacting critical points, we also propose a 2D interacting model that realizes the transition between 
$\mathcal{C}=1$ and $\mathcal{C}=2$ in Appendix C, however, analyzing the entanglement spectrum—both numerically and analytically—is substantially more challenging and is left for future work.

\section{$\mathbb{Z}$-Classified Topological Superconductors—Gapless topological physics in the three-dimensional model}
\label{sm2}

An intriguing model of $\mathbb{Z}$-classified topological superconductors of class $\rm DIII$, characterized by time-reversal symmetry $T^2=-1$ and particle-hole symmetry $C^2=1$~\cite{Altland1997PRB}, has been extensively discussed in the context of the B phase of superfluid $^3\text{He}$. In this system, fermions form spin triplet pairs, thus establishing a $p$-wave superconductor. Using a mean-field approach, we can express the BCS-pairing term by introducing appropriate order parameters:
\begin{align}
\hat H_{\text{eff}} &= \sum_{\mathbf{k},\sigma} (\epsilon_\mathbf{k} - \mu)\hat{c}_{\mathbf{k},\sigma}^{\dagger}\hat{c}_{\mathbf{k},\sigma} \nonumber\\
&+ \sum_{\mathbf{k}} \Big[\Delta_{+1}(\mathbf{k})\hat{c}_{\mathbf{k},\uparrow}^{\dagger}\hat{c}_{-\mathbf{k},\uparrow}^{\dagger} + \Delta_{+1}^*(\mathbf{k})\hat{c}_{-\mathbf{k},\uparrow}\hat{c}_{\mathbf{k},\uparrow}\Big] \nonumber\\
&+ \sum_{\mathbf{k}} \Big[\Delta_{0}(\mathbf{k})\hat{c}_{\mathbf{k},\uparrow}^{\dagger}\hat{c}_{-\mathbf{k},\downarrow}^{\dagger} + \Delta_{0}^*(\mathbf{k})\hat{c}_{-\mathbf{k},\downarrow}\hat{c}_{\mathbf{k},\uparrow}\Big] \nonumber\\
&+ \sum_{\mathbf{k}} \Big[\Delta_{-1}(\mathbf{k})\hat{c}_{\mathbf{k},\downarrow}^{\dagger}\hat{c}_{-\mathbf{k},\downarrow}^{\dagger} + \Delta_{-1}^*(\mathbf{k})\hat{c}_{-\mathbf{k},\downarrow}\hat{c}_{\mathbf{k},\downarrow}\Big] \, .
\end{align}

The $p$-wave pairing requires $\Delta_m(\mathbf{k})$ to be proportional to the spherical harmonics $\mathcal{Y}_{1,m}$, yielding:
\begin{align}
\Delta_{\pm1} &= \Delta(\pm k_x + ik_y) \, , \\
\Delta_0 &= \Delta k_z \, .
\end{align}

This continuum model represents a nontrivial topological superconductor when $\mu > 0$, as the mass term changes sign when $|\mathbf{k}| \rightarrow \infty$. To regularize this model on a square lattice, we replace $k_i$ with $\sin k_i$ and $k_i^2$ with $4\sin^2\left(\frac{k_i}{2}\right)$, where we set the lattice constant $a=1$,
\begin{align}
\hat H_{\nu=1} &=(3-\mu)\sum_{\mathbf{r}\sigma}\hat{c}_{\mathbf{r},\sigma}^{\dagger}\hat{c}_{\mathbf{r},\sigma} -t\sum_{\mathbf{r},\mathbf{e_i},\sigma}\hat{c}_{\mathbf{r},\sigma}^{\dagger}\hat{c}_{\mathbf{r}+\mathbf{e_i},\sigma}+\text{h.c.} + \sum_{\mathbf{r}} \Delta\Big(\text{i} c^\dagger_{\mathbf{r}+\mathbf{e}_x,\uparrow}c^\dagger_{\mathbf{r},\uparrow}-\text{i}c^\dagger_{\mathbf{r}-\mathbf{e}_x,\uparrow}c^\dagger_{\mathbf{r},\uparrow}+\text{h.c.}\Big) \nonumber\\
&+ \sum_{\mathbf{r}} \Delta\Big( c^\dagger_{\mathbf{r}+\mathbf{e}_y,\uparrow}c^\dagger_{\mathbf{r},\uparrow}-c^\dagger_{\mathbf{r}-\mathbf{e}_y,\uparrow}c^\dagger_{\mathbf{r},\uparrow}+\text{h.c.}\Big) + \sum_{\mathbf{r}} \Delta\Big(-\text{i} c^\dagger_{\mathbf{r}+\mathbf{e}_x,\downarrow}c^\dagger_{\mathbf{r},\downarrow}+\text{i}c^\dagger_{\mathbf{r}-\mathbf{e}_x,\downarrow}c^\dagger_{\mathbf{r},\downarrow}+\text{h.c.}\Big) \nonumber\\
&+ \sum_{\mathbf{r}} \Delta\Big( c^\dagger_{\mathbf{r}+\mathbf{e}_y,\downarrow}c^\dagger_{\mathbf{r},\downarrow}-c^\dagger_{\mathbf{r}-\mathbf{e}_y,\downarrow}c^\dagger_{\mathbf{r},\downarrow}+\text{h.c.}\Big) + \sum_{\mathbf{r}} \Delta_z\Big( -\text{i}c^\dagger_{\mathbf{r}+\mathbf{e}_z,\uparrow}c^\dagger_{\mathbf{r},\downarrow}+\text{i}c^\dagger_{\mathbf{r}-\mathbf{e}_z,\uparrow}c^\dagger_{\mathbf{r},\downarrow}+\text{h.c.}\Big) 
\end{align}
For analytical convenience, we set $\mu=3$ and $t=2\Delta=\Delta_z=1$, which establishes a topological insulator characterized by winding number $\nu=1$.  To analyze this system more efficiently, we introduce the Nambu  basis:
\begin{equation}
\hat \Psi_k^{\dagger} = (\hat{c}_{k,\uparrow}^{\dagger}, \hat{c}_{k,\downarrow}^{\dagger}, \hat{c}_{-k,\downarrow}, -\hat{c}_{-k,\uparrow}).
\end{equation}

In this representation, the effective Hamiltonian takes the elegant form:
\begin{equation}
\hat H_{\nu=1}(\mathbf{k}) = \hat \Psi_k^{\dagger} \sum_i(\sin{k_i}\alpha_i-\cos k_i\beta)\hat \Psi_k,
\end{equation}
where $\alpha_i=\sigma_x\otimes\sigma_i$ and $\beta=\sigma_z\otimes\sigma_0$ are constructed from Pauli matrices, with $\sigma_i$ ($i=x,y,z$) and $\sigma_0$ representing the identity matrix. The topological invariant of gapped phases of $\hat{H}(\mathbf{k})$ are distinguished by the winding number
\begin{equation}
\nu = \int_{\text{BZ}} \frac{d^3k}{24\pi^2}\varepsilon^{\mu\nu\rho} \text{Tr}[(q^{-1}\partial_{\mu}q)(q^{-1}\partial_{\nu}q)(q^{-1}\partial_{\rho}q)],
\end{equation}
 where $q(\mathbf{k})$ is the off-diagonal block of the flattened Hamiltonian of $H(\mathbf{k})$. From this equation, it is straightforward to construct models with higher winding number $\nu=\alpha$ starting from $\nu=1$. We simply replace $k_z$ in the original Hamiltonian with $\alpha k_z$:

\begin{equation}
\hat H_{\alpha}(\mathbf{k}) = \hat \Psi_k^{\dagger} (\sin{k_x}\alpha_x+\sin{k_y}\alpha_y+\sin{\alpha k_z}\alpha_z-(\cos{k_x}+\cos{k_y}+\cos{\alpha k_z})\beta)\hat \Psi_k 
\end{equation}
 In the corresponding lattice tight-binding model, this modification is equivalent to replacing $\mathbf{e}_z$ with $\alpha\mathbf{e}_z$ to obtain a model with higher winding number. The $\alpha$-model is a sum over all such Hamiltonian with different winding numbers. We consider the interpolations between $\nu=0,1,2$, where
 \begin{gather}
     \hat{H}^\text{3d}=\frac{1}{a+b+c}(a\hat{H}_{0}^\text{3d}+b\hat{H}_{1}^\text{3d}+c\hat{H}_{2}^\text{3d}) \, .
\end{gather} 
At $\mathbf{k}=(0,\pi,\pi)$ or $(\pi,0,\pi)$, the Hamiltonian becomes gapless when $a+c-b=0$. The nature of the phase transition depends on the relative magnitudes of the parameters. If $a>c$, this represents a trivial phase transition from $\nu=0$ to $\nu=1$. Otherwise,  if $c>a$, this represents a nontrivial phase transition from $\nu=1$ to $\nu=2$. For example, the Hamiltonian $\hat{H}_1^\text{3d}+\hat{H}_2^\text{3d}$ represents a topologically nontrivial gapless state. Additionally, there exists a tricritical point in the phase diagram at $b=2a=2c$.

\begin{figure}[tpb]
  \centering
  \includegraphics[width=1.0\linewidth]{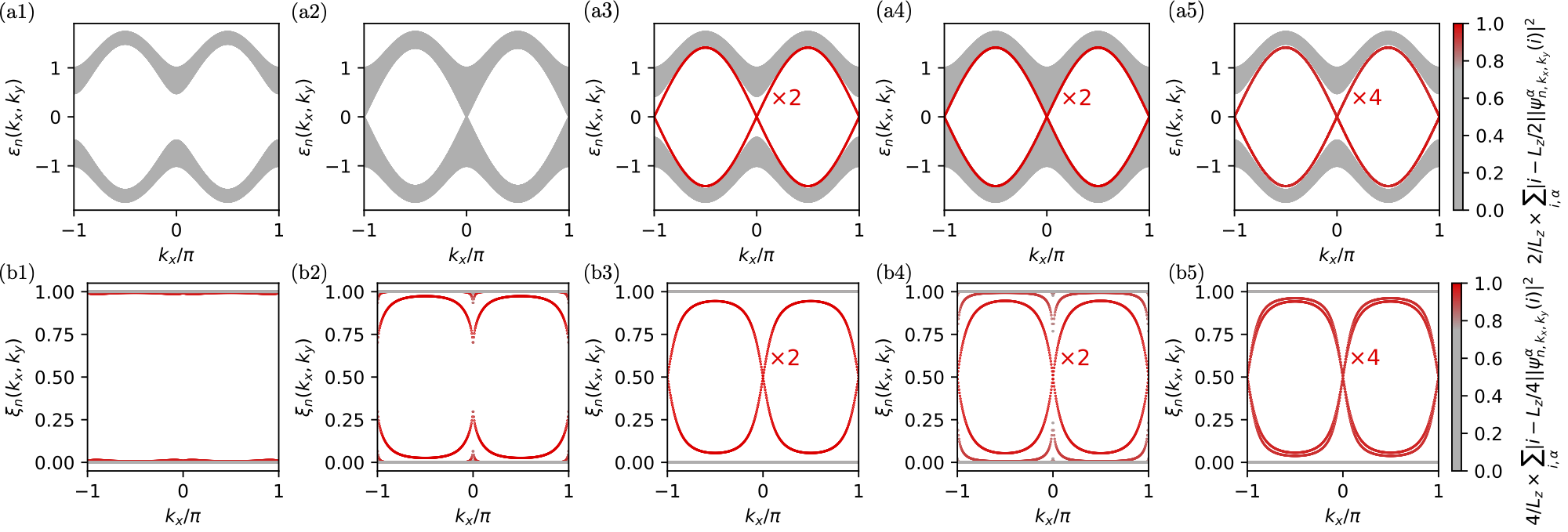}
  \caption{The energy spectrum of the 3D lattice model $(a\hat{H}^{\text{3d}}_{0}+b\hat{H}^{\text{3d}}_{1}+c\hat{H}^{\text{3d}}_{2})/(a+b+c)$ for (a1) $(a,b,c) = (4,1,1)$, (b1) $(a,b,c) = (5,6,1)$, (c1) $(a,b,c) = (1,5,1)$, (d1) $(a,b,c) = (1,6,5)$, and (e1) $(a,b,c) = (1,1,4)$, respectively. The $z$-direction is open while the $x$- and $y$-directions are periodic. (a2), (b2), (c2), (d2), and (e2) are the corresponding bipartite bulk entanglement spectrum. The entanglement cut is taken along the $z$-direction. The color coding indicates the normalized mean distance to the center along the open $z$-direction, defined as $\bar{d}_{n}(k_x,k_y) \equiv 2/L_z \sum_{i,\alpha} |i-L_z/2| \times |\psi_{n,k_x,k_y}^{\alpha}(i)|^{2}$, where $i$ is the coordinate of the unit cell in the $z$-direction. In this scheme, a value of $\bar{d}_{n}(k_x,k_y)$ close to $0$ (gray) represents a bulk state, while a value close to $1$ (red) indicates an edge mode localized at either the $z=0$ or $z=L_z$ boundary. We adopt this metric instead of the normalized mean position as used in Fig.~\ref{fig:2d_gapped} to ensure readability. Specifically, in the actual calculations, we found that the two degenerate states (marked by red colored ``$\times2$'') consist of one mode from the $z = 0$ surface and another from the $z = L_z$ surface. However, since they are degenerate, numerical diagonalization typically yields two orthogonal superpositions of these two states. For such superposed states, a simple ``mean position'' calculation might yield a value near the system center, making them indistinguishable from bulk states. In contrast, the ``mean distance'' metric correctly identifies these superpositions as localized edge states regardless of the superposition. The system size is $L_{x} = L_{y} = 400$ and $L_{z} = 40$. For simplicity, we have chosen $k_{y} = \pi - k_{x}$ in our calculations.}
  \label{fig:3d_spectra}
\end{figure}

To investigate edge modes in this 3D model, we express $\hat{H}_{\alpha}^{\text{3d}}$ via real-space operators $\hat{c}_{\bar{\mathbf{k}},n,\sigma} = \frac{1}{\sqrt{L_{z}}} \sum_{k_{z}} \text{e}^{-\text{i} n k_{z}} \hat{c}_{\mathbf{k},\sigma}^{}$ [$\bar{\mathbf{k}} \equiv (k_{x}, k_{y})$ and $\sigma = \{ \uparrow, \downarrow \}$]
\begin{equation}
    \hat{H}_{\alpha}^{\text{3d}} = \frac{1}{2} \sum_{\bar{\mathbf{k}}} \sum_{n,m} \hat{\Psi}_{\bar{\mathbf{k}},n}^{\dagger} T_{n,m}^{(\alpha)}(\bar{\mathbf{k}}) \hat{\Psi}_{\bar{\mathbf{k}},m}^{} + \text{h.c.}
\end{equation}
where $\hat{\Psi}_{\bar{\mathbf{k}},n}^{\dagger} \equiv \begin{pmatrix} \hat{c}_{\bar{\mathbf{k}},n,\uparrow}^{\dagger} & \hat{c}_{\bar{\mathbf{k}},n,\downarrow}^{\dagger} & \hat{c}_{-\bar{\mathbf{k}},n,\downarrow}^{}  & - \hat{c}_{-\bar{\mathbf{k}},n,\uparrow}^{} \end{pmatrix}$ and $T_{n,m}^{(\alpha)}(\bar{\mathbf{k}}) \equiv \delta_{n,m} \big( \sin(k_{x}) \alpha_{x} + \sin(k_{y}) \alpha_{y} - \cos(k_{x}) \beta - \cos(k_{y}) \beta \big) - \delta_{n,m+\alpha} \big( \text{i} \alpha_{z} + \beta \big)$. 
Here, we consider open boundaries along the $z$-direction and use $(k_{x}, k_{y})$ as good quantum numbers.

The upper panel of Fig.~\ref{fig:3d_spectra} shows the energy spectra for representative parameters $(a, b, c) = (4, 1, 1)$, $(5, 6, 1)$, $(1, 5, 1)$, $(1, 6, 5)$, and $(1, 1, 4)$, respectively.
For the gapped models with $\nu = 1$ (c1) and $\nu = 2$ (e1), we can observe the gapless edge modes. 
Importantly, we see no edge modes at the trivial critical point [see Fig.~\ref{fig:3d_spectra} (b1)], while clear edge modes exist at the topologically nontrivial critical point [see Fig.~\ref{fig:3d_spectra} (d1)]. 
The corresponding bulk entanglement spectra shown in the lower panel of Fig.~\ref{fig:3d_spectra} exhibit similar behaviors that confirms the bulk-boundary correspondence proposed in our work for 3D cases.

\twocolumngrid
\let\oldaddcontentsline\addcontentsline
\renewcommand{\addcontentsline}[3]{}
\bibliography{main.bib}
\let\addcontentsline\oldaddcontentsline

\end{document}